\documentclass[11pt, a4paper]{article}
\usepackage{fullpage}

\usepackage{belov_paper}

\title{Quantum Lower Bounds for Tripartite Versions of the Hidden Shift and the Set Equality Problems}
\author{Aleksandrs Belovs\thanks{Faculty of Computing, University of Latvia}
\and
Ansis Rosmanis\thanks{Centre for Quantum Technologies, National University of Singapore}}
\date{}

\newcommand{\gG}{\mathfrak{G}}
\newcommand{\gS}{\mathfrak{S}}

\begin{document}

\maketitle

\begin{abstract}
In this paper, we study quantum query complexity of the following rather natural tripartite generalisations (in the spirit of the 3-sum problem) of the hidden shift and the set equality problems, which we call the 3-shift-sum and the 3-matching-sum problems.

The 3-shift-sum problem is as follows: given a table of $3\times n$ elements, is it possible to circularly shift its rows so that the sum of the elements in each column becomes zero?  It is promised that, if this is not the case, then no 3 elements in the table sum up to zero.
The 3-matching-sum problem is defined similarly, but it is allowed to arbitrarily permute elements within each row.
For these problems, we prove lower bounds of $\Omega(n^{1/3})$ and $\Omega(\sqrt n)$, respectively.  The second lower bound is tight.

The lower bounds are proven by a novel application of the dual learning graph framework and by using representation-theoretic tools from~\cite{belovs:setEquality}.
\end{abstract}

\newcommand{\mm}{{\textsc{m}}}
\newcommand{\sss}{{\textsc{s}}}
\newcommand{\qq}{{\textsc{q}}}
\newcommand{\Adv}{\mathop{\mathrm{ADV}^{\pm}}}

\newcommand{\smap}[1]{\stackrel{#1}{\longmapsto}}

\section{Introduction}

One of the starting points of this paper was the following problem, posed by Aaronson and Ambainis~\cite{aaronson:forrelation}: construct a partial Boolean function with polylogarithmic quantum query complexity but whose randomised query complexity is $\omega(\sqrt n)$, where $n$ is the number of input variables.
There are relatively many functions known with the required quantum query complexity and randomised query complexity $\Theta(\sqrt{n})$.
For instance, one can take the forrelation problem of~\cite{aaronson:forrelation} with quantum query complexity 1, or the better-known hidden subgroup problem~\cite{ettinger:hspQuery}.
However, no function with polylogarithmic quantum query complexity and randomised query complexity $\omega(\sqrt n)$ is known.
As shown in~\cite{ben-david:super-grover, aaronson:cheatSheets}, such a function would also yield a larger than $5/2$-power separation between quantum and randomised query complexities for \emph{total} Boolean functions.

Aaronson and Ambainis proposed a candidate function, which they call the $k$-fold forrelation.
It has a very simple quantum $O(1)$-query algorithm, but it seems hard to lower bound its randomised query complexity.
However, it is also possible to go in the opposite direction: find a function whose randomised query complexity is $\omega(\sqrt n)$, and construct an efficient quantum algorithm computing this function.  
A potential candidate might be a modification of a function already known to be easy quantumly, preserving the hope the modification is still easy. 

One particularly neat starting function, in our opinion, is the following \emph{hidden shift problem}.
Given two stings $x,y\in[q]^{n}$, the task is to distinguish two cases:
in the positive case, $x$ is a circular shift of $y$;
in the negative case, all the input variables in $x$ and $y$ are distinct.
This problem is equivalent to the hidden subgroup problem in the dihedral group~\cite{kuperberg:dihedral}, and its quantum query complexity is logarithmic.  It is also easy to see that its randomised query complexity is $\Theta(\sqrt n)$.

In this paper we consider the following modification, which we call the \emph{3-shift-sum} problem.
We are given an input string $x\in[q]^{3n}$, which we treat as a $3\times n$ table.  In the positive case, it is possible to circularly shifts the rows of the table so that the sum of the elements in each column becomes divisible by $q$.  In the negative case, no matter how we shift the rows, there is no column with the sum of its elements divisible by $q$.
(In other words, there is no three elements, one from each row, whose sum is divisible by $q$.)
This is a natural amalgamation of the hidden shift and the 3-sum problem, both studied quantumly.

It is easy to see that the randomised query complexity of this problem is $\Theta(n^{2/3})$.
This raises the question of what its quantum query complexity is.
Our first result is a simple proof that, unlike the hidden shift problem, the quantum query complexity of the 3-shift-sum problem is polynomial: $\Omega(n^{1/3})$.
Thus, the 3-shift-sum problem fails to provide the desired separation.

Similarly as the 3-shift-sum problem is a tripartite version of the hidden shift problem, the \emph{3-matching-sum} problem is a tripartite version of the \emph{set equality} problem.
In the set equality problem, the negative inputs are as in the hidden shift problem, but in a positive input, $y$ is an arbitrary permutation of $x$, not necessary a circular shift.
Unlike the hidden shift problem, the set equality problem has polynomial quantum query complexity: $\Theta(n^{1/3})$~\cite{shi:collisionLowerOriginal, zhandry:setEquality, belovs:setEquality}.
In our tripartite version of it, the negative inputs are the same as in the 3-shift-sum problem, but for a positive input, there exists an arbitrary permutation of the elements within each row such that the sum of each column becomes divisible by $q$.
Our second result is a complete characterisation of the quantum query complexity of this problem: it is $\Theta(\sqrt n)$.

\paragraph{Techniques}
Our main tool is the framework of dual learning graphs, which is ``compiled'' to the adversary lower bound.

The first version of the adversary method was developed by Ambainis~\cite{ambainis:adv}.  This version, later known as the positive-weighted adversary, is easy to use, and it has found many applications, but it is also subject to some limitations: the certificate complexity~\cite{spalek:advEquivalent, zhang:advPower} and the property testing~\cite{hoyer:advNegative} barriers.
The property testing barrier, which is relevant to our problems, states that, if any positive input differs from any negative input on at least $\eps$ fraction of the input variables, the positive-weighted adversary fails to prove a lower bound better than $\Omega(1/\eps)$.
In most cases $\eps=\Omega(1)$, thus this only gives a trivial lower bound.

The next version of the bound, the negative-weighted adversary~\cite{hoyer:advNegative}, is known to be tight~\cite{reichardt:spanPrograms}, but it is also harder to apply.
An application of the bound to the $k$-sum problem was obtained in~\cite{spalek:kSumLower}.  This result was later stated in the framework of dual learning graphs~\cite{belovs:onThePower}, which we are about to describe.

Learning graphs is a model of computation introduced in~\cite{belovs:learning, belovs:phd}.
They are most naturally stated in terms of certificate structures, which describe where 1-certificates can be located in a positive input.
Learning graphs capture quantum query complexity of certificate structures in the following sense.  Let $L$ be the learning graph complexity of a certificate structure $\cC$.
First, for \emph{any} function with certificate structure $\cC$, there exists a quantum algorithm solving it in $O(L)$ queries.  
Second, there exists \emph{some} function with certificate structure $\cC$ and quantum query complexity $\Omega(L)$.
In general these functions are rather contrived, yet one example of them being natural are the following sum problems.
A sum problem is a \emph{total} function parametrised by a family $\cS$ of $O(1)$-sized subsets of $[n]$.
The task, given an input string $x\in[q]^n$, is to detect whether there exists $S\in\cS$ such that $\sum_{i\in S} x_i$ is divisible by $q$.
Note that our problems do not fall into this category, because every positive input is promised to have many such subsets.

While dual learning graphs give tight lower bounds for all of the above sum problems,
in general, of course, they do not give lower bounds for all problems with a given certificate structure.
For example, the learning graph complexity of the certificate structure corresponding to the hidden shift problem is $\Theta(n^{1/3})$, whereas its quantum query complexity is logarithmic.
What about the 3-shift-sum problem?  It turns out that dual learning graphs are still of help here, but in a slightly different way.
The learning graph complexity of the corresponding certificate structure is $\Theta(\sqrt n)$, yet we do not know whether it can be converted into a quantum query lower bound.
However, a dual learning graph for a \emph{different} certificate structure can be converted into, albeit not tight, but still a polynomial lower bound.
This shows that dual learning graphs are more versatile than we thought.

Another interesting feature of our result is that it might be the simplest constructed example of the adversary bound surpassing the property testing barrier.
Examples of the negative-weighted adversary breaking the certificate complexity and the property testing barriers were already obtained in~\cite{hoyer:advNegative}.  
But~\cite{hoyer:advNegative} did not cover the most interesting regime $\eps = \Omega(1)$ of the property testing barrier.
The sum problems of~\cite{belovs:onThePower} are relatively simple examples of overcoming the certificate complexity barrier.
An example for the $\eps=\Omega(1)$ regime of the property testing barrier was constructed in~\cite{belovs:setEquality}, but the construction is quite technical.
Our result gives a similar example by much simple means, comparable to that of~\cite{belovs:onThePower}.

Concerning the 3-matching-sum problem, our lower bound is an application of the technique developed for the set equality problem~\cite{belovs:setEquality}.  It is based on the representation theory of the symmetric group.  Surprisingly, the technique can be used for the 3-matching-sum problem with essentially no modifications:
our proof uses representation theory to a minimal extent, and mostly follows from combinatorial estimates involving the dual learning graph.
This indicates that our technique has a potential to be used in proving lower bounds for other symmetric problems.

\paragraph{Results in property testing}
In the property testing model, one is given some property (a set of positive inputs), and the task is to distinguish whether the input possesses the property, or is $\eps$-far, in the relative Hamming distance, from any input that has the property.

Overcoming the property testing barrier automatically gives a lower bound for a property testing problem---that of testing whether the input is positive.  But it is not always the most natural way to state the problem.
We give an example of a lower bound for a problem that is most naturally stated in the setting of property testing.

The 3-shift-sum problem, as formulated above, must have relatively large $q$ for the problem to be interesting.  For instance, it is easy to see that for $q=2$ there are almost no negative inputs.  In our lower bound, we require that $q=\Omega(n^3)$.
But it is possible to formulate a version of the problem that is interesting even when the input alphabet is Boolean.
Define the set of positive inputs as before, and define the set of negative inputs as being at relative Hamming distance at least, say, $1/7$ to it.
We prove a lower bound of $\Omega(n^{1/3})$ also for this version of the problem.

Although there is quite a number of quantum algorithms for property testing problems, there are not so many quantum lower bounds known. (An interested reader might consult a recent survey~\cite{montanaro:quantumProperyTest} for more information on the topic.)
One of the main reasons, of course, is the property testing barrier for the positive-weighted adversary.
Up to our knowledge, our result is the first property testing lower bound proven using the adversary method, which answers the problem mentioned in~\cite{montanaro:quantumProperyTest}.
This shows yet another area of applications of dual learning graphs.

\section{Preliminaries}

For positive integers $m$ and $\ell \geq m$, let $[m]$ denote the set $\{1,2,...,m\}$ and $[m..\ell]$ denote the set $\{m,m+1,\ldots,\ell\}$.
For $P$ a predicate, we use $1_P$ to denote the variable that is 1 is $P$ is true, and 0 otherwise.

For an $\cI\times\cJ$-matrix $A$, $i\in\cI$, and $j\in\cJ$, we denote by $A\elem[i,j]$ its $(i,j)$-th entry.
For $\cI'\subseteq\cI$ and $\cJ'\subseteq\cJ$, $A\elem[\cI', \cJ']$ denotes the corresponding submatrix.
We use similar notation also for vectors.
Next, $\norm|\cdot|$ denotes the spectral norm (the largest singular value), and $\circ$ denotes the Hadamard (i.e., entry-wise) product of matrices.
We often identify projectors with the spaces they project onto.

\subsection{Adversary Bound}
For background on quantum query complexity the reader may refer to~\cite{buhrman:querySurvey}.  
In the paper, we only require the knowledge of the (negative-weighted) adversary bound for decision problems, which we are about to define.

Let $f\colon \cD\to\bool$ with $\cD\subseteq[q]^n$.
An {\em adversary matrix} for $f$ is a non-zero $f^{-1}(1)\times f^{-1}(0)$-matrix $\Gamma$.
 For any $j\in[n]$, the $f^{-1}(1)\times f^{-1}(0)$-matrix $\Delta_j$ is defined by
\begin{equation}
\label{eqn:Delta}
 \Delta_j[\![x,y]\!] = \begin{cases}0,&\text{if $x_j=y_j$;}\\1,&\text{if $x_j\neq y_j$.}\end{cases}
\end{equation}

\begin{thm}[Adversary bound~\cite{hoyer:advNegative,lee:stateConversion,spalek:kSumLower}]
\label{thm:adversary}
In the above notation, the quantum query complexity of the function $f$ is $\Theta\sA[\Adv(f)]$, where $\Adv(f)$ is the optimal value
of the semi-definite program
\begin{subequations}
\label{eqn:adv}
\begin{alignat}{3}
 &{\mbox{\rm maximise }} &\quad& \norm|\Gamma| \\ 
 &{\mbox{\rm subject to }} && \norm|\Delta_j\circ\Gamma|\leq 1\qquad \mbox{ for all }j\in[m] \label{eqn:advCondition},
\end{alignat}
\end{subequations}
where the maximisation is over all adversary matrices $\Gamma$ for $f$.
\end{thm}
We can choose any adversary matrix $\Gamma$ and scale it so that the condition
$\norm|\Delta_j\circ\Gamma|\leq 1$ holds.  Thus, we often use the condition
$\norm|\Delta_j\circ\Gamma|= \mathrm{O}(1)$ instead of $\norm|\Delta_j\circ\Gamma|\leq 1$.

Working with the matrix $\Delta_j\circ\Gamma$ might be cumbersome, so we do the following transformation instead.  We write $\Gamma\smap{\Delta_j} \Gamma'$ if $\Gamma\circ\Delta_j = \Gamma'\circ\Delta_j$.  In other words, we modify the entries of $\Gamma$ with $x_j = y_j$.  
Now, from the fact~\cite{lee:stateConversion} that
$
\gamma_2(\Delta_j) = \max_B\big\{\norm|\Delta_j\circ B|:\norm|B|\leq 1\big\}\leq 2
$,
we conclude that $\|\Delta_j\circ \Gamma\| \le 2\|\Gamma_j\|$, hence we can replace $\Delta_j\circ\Gamma$ with $\Gamma'$ in~\rf{eqn:advCondition}.

It is sometimes convenient~\cite{spalek:kSumLower} to allow several rows or columns corresponding to the same input $x$.  We add labels to distinguish different rows corresponding to the same input.

\subsection{Certificate Structures and Dual Learning Graphs}

Let $f\colon \cD\to\bool$ be a function with domain $\cD\subseteq [q]^n$.
For $x\in f^{-1}(1)$, a \emph{certificate} for $x$ is a subset $S\subseteq[n]$ such that $f(z)=1$ for all $z\in\cD$ satisfying $x_i = y_i$ for all $i\in S$.
A \emph{certificate structure} $\cC$ is a collection of non-empty subsets of $2^{[n]}$.
We say that $f$ \emph{has} certificate structure $\cC$ if, for every $x\in f^{-1}(1)$, there exists $\cM\in\cC$ such that every $S\in\cM$ is a certificate for $x$.
It is natural to assume that all $\cM\in\cC$ are upward closed.

There are two formulations of the learning graph complexity: primal and dual.  For the purposes of this paper, it is enough to state the dual one.
A \emph{dual learning graph} for a certificate structure $\cC$ is a feasible solution to the following optimisation problem:
\begin{subequations}
\label{eqn:learningDual}
\begin{alignat}{3}
 &{\mbox{\rm maximise}} &\;& \sqrt{\sum\nolimits_{\cM\in \cC} \alpha(\cM, \emptyset)^2} \label{eqn:alphaObjective} \\ 
 &{\mbox{\rm subject to}} && \sum_{\cM\in \cC} \sA[\alpha(\cM, S) - \alpha(\cM, S\cup\{j\})]^2\le 1 &\quad&\text{\rm for all $S\subseteq[n]$ and $j\in[n]\setminus S$;} \label{eqn:alphaOne} \\
 &&& \alpha(\cM, S) = 0 && \text{\rm whenever $S\in\cM$;} \label{eqn:alphaZero}  \\
 &&& \alpha(\cM, S)\in\bR && \text{\rm for all $\cM\in \cC$, $S\subseteq[n]$, and $S\notin\cM$.} \label{eqn:alphaDef}
\end{alignat}
\end{subequations}
The optimal value of this optimisation problem is called the learning graph complexity of $\cC$.

We call a solution to the dual learning graph for $\cC$ any mapping $\alpha(\cM, S)$ satisfying~\rf{eqn:alphaDef}, where we implicitly assume~\rf{eqn:alphaZero}.  A solution is feasible if it satisfies~\rf{eqn:alphaOne}.
It is easy to see that any optimal solution $\alpha(\cM, S)$ to~\rf{eqn:learningDual} is entry-wise non-negative and non-increasing in $S$.  We will implicitly assume that any feasible solution satisfies these requirements.

Inspired by this optimisation problem, we define the norm of a solution $\alpha$ as
\[
\|\alpha\| = \max_{S\subseteq [n]} \sqrt{\sum\nolimits_{\cM\in \cC} \alpha(\cM, S)^2 }.
\]
It satisfies the usual axioms of a norm, although we will not use this fact.  For $j\in[n]$, we define an operation $\partial_j$ given by
\[
\partial_j \alpha(\cM, S) =
\begin{cases}
\alpha(\cM, S) - \alpha(\cM, S\cup\{j\}), &\text{if $j\notin S$;}\\
0, & \text{if $j\in S$.}
\end{cases}
\]
If $\alpha$ is a solution to the dual learning graph, so is $\partial_j\alpha$.  Condition~\rf{eqn:alphaOne} can be restated as $\|\partial_j \alpha\|\le 1$ for all $j\in[n]$.  If $\alpha(\cM, S)$ is non-increasing in $S$, the objective value~\rf{eqn:alphaObjective} is given by $\|\alpha\|$.

Dual learning graphs have close connection to adversary matrices, which we discuss in \rf{sec:basic}.

\subsection{Representation Theory}
\label{sec:representation}

In this section, we introduce basic notions from the representation theory of finite groups with special emphasis on the symmetric group.  For more background, the reader may refer to~\cite{curtis:representationTheory, serre:representation} for general theory, and to~\cite{james:symmetricGroup, sagan:symmetricGroup} for the special case of the symmetric group.

Assume $G$ is a finite group.  
The \emph{group algebra} $\bC G$ is the complex vector space with the elements of $G$ forming an orthonormal basis, where the multiplication law of $G$ is extended to $\bC G$ by linearity.
A (left) \emph{$G$-module}, also called a \emph{representation} of $G$, is a complex vector space $V$ with the left multiplication operation by the elements of $\bC G$ satisfying the usual associativity and distributivity conditions.  
We can treat elements of $\bC G$ as linear operators acting on $V$.
%

A \emph{$G$-morphism} (or just morphism, if $G$ is clear from the context) between two $G$-modules $V$ and $W$ is a linear operator $\theta\colon V\to W$ that commutes with all $\alpha\in\bC G$: $\theta\alpha = \alpha\theta$, where the first $\alpha$ acts on $V$ and the second one on $W$.

A $G$-module is called \emph{irreducible} (or just irrep for irreducible representation) if it does not contain a non-trivial $G$-submodule.
For any $G$-module $V$, one can define its \emph{canonical} decomposition into the direct sum of \emph{isotypic} subspaces, each spanned by all copies of a fixed irrep in $V$.
Different isotypic subspaces in this decomposition are orthogonal.  
If an isotypic subspace contains at least one copy of the irrep, we say that $V$ {\em uses} this irrep.

If $G$ and $H$ are finite groups, then the irreducible $G\times H$-modules are of the form $V\otimes W$ where $V$ is an irreducible $G$-module and $W$ is an irreducible $H$-module.  And the corresponding group action is given by $(g,h)(v\otimes w) = gv\otimes hw$, with $g\in G$, $h\in H$, $v\in V$, and $w\in W$, which is extended by linearity.

We use the following results:
\begin{lem}[Schur's Lemma]
\label{lem:schur}
Assume $\theta\colon V\to W$ is a morphism between two irreducible $G$-modules $V$ and $W$. 
Then, $\theta=0$ if $V$ and $W$ are non-isomorphic; otherwise, $\theta$ is uniquely determined up to a scalar multiplier.
\end{lem}
\begin{lem}[\cite{belovs:setEquality}]
\label{lem:singularirrep}
Let $\theta\colon V\to W$ be a morphism between two $G$-modules $V$ and $W$.  Then, there exists an irrep in $V$ all consisting of right-singular vectors of $\theta$ of singular value $\|\theta\|$. (We call such right-singular vectors principal.)
\end{lem}

Let $\gS_L$ denote the \emph{symmetric group} on a finite set $L$, that is, the group with the permutations of $L$ as elements, and composition as the group operation.  If $m$ is a positive integer, $\gS_m$ denotes the isomorphism class of the symmetric groups $\gS_L$ with $|L|=m$.
Representation theory of $\gS_m$ is closely related to \emph{Young diagrams}, defined as follows.

A \emph{partition} $\lambda$ of an integer $m$ is a non-increasing sequence $(\lambda_1,\dots,\lambda_t)$ of positive integers satisfying $\lambda_1+\dots+\lambda_t = m$.  
A partition $\lambda = (\lambda_1,\dots,\lambda_t)$ is often represented in the form of a Young diagram that consists, from top to bottom, of rows of $\lambda_1,\lambda_2,\dots,\lambda_t$ boxes aligned by the left side.
We say that a partition \emph{has $k$ boxes below the first row} if $\lambda_1 = m-k$.
For each partition $\lambda$ of $m$, there exists an irreducible $\gS_m$-module $S^{\lambda}$, called the \emph{Specht module}.  All these modules are pairwise non-isomorphic, and give a complete list of all the irreps of $\gS_m$.

\section{Formulation of the Problems and Easy Observations}
\label{sec:formulations}

In this section, we formulate the 3-shift-sum problem and define the closely related 3-matching-sum problem.
We also sketch proofs of few simple observations about these problems.

Both the 3-shift-sum and the 3-matching-sum are partial Boolean functions defined on $[q]^{3n}$, with $q$ and $n$ positive integers.  The $3n$ input variables are divided into three groups $A=[1..n]$, $B=[n+1..2n]$, and $C=[2n+1..3n]$.
A \emph{3-dimensional matching} is a partition $\mu$ of the set $[3n]$ into $n$ triples, $\mu = \{T_1,\dots,T_n\}$, such that $|T_i\cap A| = |T_i\cap B| = |T_i\cap C| = 1$ for all $i$.  This is a natural generalisation of the usual (2-dimensional) matching between sets $A$ and $B$.
We denote the set of 3-dimensional matchings by $M_\mm$ (we omit $n$, assuming its value is clear from the context).
We consider a special type of 3-dimensional matchings, we call \emph{3-shifts}.  A 3-shift is a matching $\mu = \{T_1,\dots,T_n\}$ such that there exist two numbers $b,c\in [n]$ such that $T_i = \sfigA{i,\; n+1+(i+b \bmod n),\; 2n+1+(i+c \bmod n)}$ for all $i$.  We denote the set of 3-shifts by $M_\sss$.

We define the \emph{3-shift-sum} and the \emph{3-matching-sum} problems as follows.  
Let $M_\qq$ stand for $M_\sss$ in 3-shift-sum and for $M_\mm$ in 3-matching-sum. 
In a positive input $x\in[q]^{3n}$, there exists $\mu\in M_\qq$ such that $x_a + x_b + x_c$ is divisible by $q$ for every triple $\{a,b,c\}\in\mu$.
We say that $x$ is \emph{of the form} $\mu$ in this case.
In a negative input $y\in [q]^{3n}$, we have $y_a + y_b + y_c \not\equiv 0 \pmod q$ for any choice of $a\in A$, $b\in B$, and $c\in C$.
The task is to determine whether the input is positive or negative, provided that one of the two options holds.
Since 3-shift-sum is a special case of 3-matching-sum, the latter is a harder problem.

\paragraph{Randomised and Quantum Complexity}
Let us describe what we can immediately say about quantum and randomised query complexities of these problems.  Neither result will be relevant later in the paper.

\begin{prp}
\label{prp:quantumAlgorithm}
The quantum query complexity of the 3-shift-sum and the 3-matching-sum problems is $O(\sqrt n)$.
\end{prp}

\pfstart[Proof sketch]
Consider a positive input $x$, and let $\mu\in M_\qq$ be its form.
Take random subsets $A'\subseteq A$ and $B'\subseteq B$ of size approximately $\sqrt n$, and query all the variables in $A'\cup B'$.
With high probability, there exists $T\in\mu$ that intersects both $A'$ and $B'$.  Now use Grover's search to find an element $c\in C$ satisfying $x_a + x_b + x_c \equiv 0\pmod q$ for some $a\in A'$ and $b\in B'$.
\pfend

\begin{prp}
\label{prp:randomised}
The randomised query complexity of the 3-shift-sum and the 3-matching-sum problems is $\Theta(n^{2/3})$.
\end{prp}

\pfstart[Proof sketch]
Let us start with the upper bound.
Let $x$ be a positive input of the form $\mu = \{T_1,\dots,T_n\}$.
Query random subsets $A'\subseteq A$, $B'\subseteq B$, and $C'\subseteq C$ of size approximately $n^{2/3}$.
With high probability there exists $T\in M_{\qq}$ that intersects all $A'$, $B'$, and $C'$.

For the lower bound, we assume that $q\gg n^3$, so that almost all inputs are negative.
Consider two probability distributions $\cP$ and $\cU$ on the inputs defined as follows.
The distribution $\cU$ is uniform over all input strings in $[q]^{3n}$.
Since almost all inputs are negative, $\cU$ is close to the distribution $\cN$ over all negative inputs.  
In the distribution $\cP$, a 3-dimensional matching $\mu\in M_\qq$ is chosen uniformly at random.
Then, a string in $[q]^{2n}$ is chosen uniformly at random, and uniquely extended to a positive input of the form $\mu$.

Let $S\subseteq[3n]$.
If $|S|\ll n^{2/3}$, the probability $S$ contains a triple from $\mu$ as a subset is negligible.
Conditioned on this not happening, the distribution $\cP$ is completely indistinguishable from $\cU$ given the values of the variables in $S$.
Hence, no randomised algorithm using considerably fewer than $n^{2/3}$ queries can distinguish $\cP$ and $\cU$ with non-negligible probability.
Since $\cU$ is close to $\cN$, we get that any randomised algorithm solving the 3-shift-sum or the 3-matching-sum problem with bounded error uses $\Omega(n^{2/3})$ queries.
\pfend

\paragraph{Certificate Structures}
It is easy to describe the certificate structures $\cC_\sss$ and $\cC_\mm$ of the 3-shift-sum and the 3-matching-sum problems.  For each $\mu\in M_\qq$, there is a corresponding $\cM_\mu \in \cC_\qq$ obtained as follows: a subset $S\subseteq[n]$ is in $\cM_\mu$ if and only if there exists a triple $T\in \mu$ satisfying $T\subseteq S$.

The lower bound from the following proposition will be our main source of inspiration when constructing adversary bounds later in the paper.

\begin{prp}
\label{prp:learningBoth}
The learning graph complexity of the certificate structures $\cC_\sss$ and $\cC_\mm$ is $\Theta(\sqrt n)$.
\end{prp}

\pfstart
The upper bound is similar to \rf{prp:quantumAlgorithm}, and we omit the proof.  The upper bound is stated here for completeness, and we do not use it further in the paper.

Let us prove the lower bound.  For that we have to construct a feasible solution to the dual learning graph.
For $\cM\in \cC_\qq$, define
\begin{equation}
\label{eqn:alphacMSopt}
\alpha(\cM, S) = \frac1{\sqrt{|M_\qq|}} \max\sfig{\sqrt n - |S|, 0}\qquad \text{if $S\notin \cM$},
\end{equation}
and as 0 otherwise.
It is easy to see that the objective value~\rf{eqn:alphaObjective} is $\sqrt n$, and that~\rf{eqn:alphaZero} holds.

It remains to check~\rf{eqn:alphaOne}.
Fix $S$ and $j$.
If $|S|\ge \sqrt n$, then the left-hand side of~\rf{eqn:alphaOne} is zero, so assume $|S|\le \sqrt n$.
We have the following contributions to the left-hand side of~\rf{eqn:alphaOne}:

\begin{itemize}
\item If $S\cup\{j\}\notin\cM$, then the value of $\alpha(\cM, S)$ changes by $\frac{1}{\sqrt{|M_\qq|}}$ as $|S|$ increases by 1.
\item If $\mu\in M_\qq$ is taken uniformly at random,
the probability is $O\s[(|S|/n)^2] = O(1/n)$ that $S\notin\cM_\mu$ but $S\cup\{j\}\in\cM_\mu$.
In this case, $\alpha(\cM_\mu, S)$ changes by at most $\sqrt{\frac{n}{|M_\qq|}}$.
\end{itemize}
Altogether we have:
\[
\sum_{\cM\in \cC} \sA[\alpha(\cM, S) - \alpha(\cM, S\cup\{j\})]^2
\le |M_\qq| \cdot\frac{1}{|M_\qq|} + O\s[\frac{|M_\qq|}n] \cdot \frac{n}{|M_\qq|} = O(1).
\]
Scaling down $\alpha$ by a constant factor, we get a feasible solution with objective value $\Omega(\sqrt n)$.
\pfend

\newcommand{\hGamma}{\widehat\Gamma}
\newcommand{\hG}{\widehat G}

\section{Basic Definitions}
\label{sec:basic}
In this section we introduce our basic notation, and describe a procedure of converting a solution to the dual learning graph into an adversary matrix.  
This is a general procedure from~\cite{belovs:onThePower} tailored for the special case of the 3-shift-sum and the 3-matching-sum problems.
This procedure does not immediately result in good adversary matrices for these problems, but we are able modify it in Sections~\ref{sec:3shift} and~\ref{sec:3matching} so that it works.
Let again $M_\qq$ stand for either $M_\sss$ or $M_\mm$.

\paragraph{Fourier Basis}
Let $\cH = \bC^{\bZ_q}$ and $e_0,\dots,e_{q-1}$ be the Fourier basis of $\cH$. 
Recall that it is an orthonormal basis given by $e_i\elem[j] = \frac1{\sqrt q}\omega^{ij}$, where $\omega = \ee^{2\pi\ii/q}$.
For $m$ a positive integer, the Fourier basis of $\cH^{\otimes m}$ is given by tensor products $e_{a_1}\otimes\cdots\otimes e_{a_m}$.  A component $e_{a_i}$ in this tensor product is called non-zero if $a_i\ne 0$.  The weight of the Fourier basis element is the number of non-zero components.

We define two projectors in $\cH$:
\[
\Pi_0 = e_0e_0^* \qqand \Pi_1 = I - \Pi_0 = \sum_{i=1}^{q-1} e_ie_i^*.
\]
All the entries of $\Pi_0$ are equal to $1/q$.
An important relation is 
\begin{equation}
\label{eqn:EDelta}
\Pi_0\smap{\Delta}\Pi_0 
\qqand
\Pi_1\smap{\Delta} - \Pi_0,
\end{equation}
where $\Delta$ is as in~\rf{eqn:Delta} and acts on the sole variable.
For two sets $R\subseteq T$, we define a projector $\Pi^T_R$ in the space $\cH^T$ by
\[
\Pi^T_R = \bigotimes\nolimits_{j\in T} \Pi_{1_{j\in R}}.
\]
As $R$ ranges over all subsets of $T$, this gives an orthogonal decomposition of $\cH^T$.
By~\rf{eqn:EDelta}:
\begin{equation}
\label{eqn:PiDelta}
\text{$\Pi^{T}_R \smap{\Delta_j} \Pi^T_R$\quad if\quad $j\notin R$}
\qqand
\text{$\Pi^{T}_R \smap{\Delta_j} -\Pi^{T}_{R\setminus\{j\}}$\quad if\quad $j\in R$}.
\end{equation}

If $\cA$ is a collection of subsets of $T$, we can define projector $\Pi^T_{\cA} = \sum_{R\in\cA}\Pi^T_R$.  We clearly have $\Pi_{\cA}\Pi_{\cB} = \Pi_{\cA\cap\cB}$. We will use this construction only for some special cases, in particular, for a positive integer $k$, we define $\Pi^T_k = \sum_{R\subseteq T, |R|=k} \Pi^T_R$.

\paragraph{Basic Operators}
Let $\mu = \{T_1,\dots,T_n\}$ be a 3-dimensional matching.
Let $\cP^\mu$ denote the set of positive inputs of form $\mu$.
We use $\cP$ for the set of pairs $(\mu, x)$ with $\mu\in M_\qq$ and $x\in \cP^\mu$.
Think of $\cP$ as the set of positive inputs with additional labels so that some inputs $x$ can appear multiple times.
We use $\cN$ for the set of negative inputs, and $\cU = [q]^{3n}$ for the set of all strings.
Similarly to the proof of \rf{prp:randomised}, $\cU$ will be close to $\cN$, and we use the former as a proxy for the latter.

Now assume $T$ is a triple of elements.  Think of it as an element of a 3-dimensional matching $\mu$.
Denote
\begin{equation}
\label{eqn:P}
P^T=\sfig{(a,b,c)\in[q]^T\mid a+b+c \equiv 0\pmod q}.
\end{equation}
Thus, $\cP^\mu$ is the Cartesian product $\prod_{T\in\mu} P^T$.
For $R\subseteq T$, define 
\begin{equation}
\label{eqn:PsiTR}
\Psi^T_R = \sqrt{q}\; \Pi^{T}_R\, \elem[P^T, {[q]^T}],
\end{equation}
where the factor $\sqrt q$ is introduced to account for the reduced number of rows.
For $S\subseteq [3n]$, let
\[
\Psi^{\mu}_S = \bigotimes_{T\in\mu}\nolimits \Psi^T_{S\cap T} 
= q^{n/2}\, \Pi^{[3n]}_S\elem[\cP^\mu, \cU].
\]
As for $\Pi^T_\cA$, we will use $\Psi^\mu_\cA = \sum_{S\in\cA} \Psi^\mu_S$ for a family $\cA$ of subsets of $[3n]$.  Again, $\Psi^\mu_\cA \Pi^{[3n]}_\cB = \Psi^\mu_{\cA\cap \cB}$.

Using~\rf{eqn:PiDelta}, we have
\begin{equation}
\label{eqn:PsiDelta}
\text{$\Psi^\mu_S \smap{\Delta_j} \Psi_S^\mu$\quad if\quad $j\notin S$}
\qqand
\text{$\Psi^\mu_S \smap{\Delta_j} -\Psi_{S\setminus\{j\}}^\mu$\quad if\quad $j\in S$}.
\end{equation}

\paragraph{From Dual Learning Graphs to Adversary Matrices}

Now we explain how to convert a solution $\alpha$ to the dual learning graph~\rf{eqn:learningDual} into a $\cP\times\cU$-matrix $G(\alpha)$.
In ~\cite{belovs:onThePower}, the adversary matrix $\Gamma$ was obtained by restricting $\Gamma = G(\alpha)\elem[\cP, \cN]$.  It is convenient to allow all the columns corresponding to $\cU$, and restrict them to $\cN$ only at the very end.

If $\mu\in M_\qq$, let us for brevity write $\alpha(\mu,S)$ for $\alpha(\cM_\mu, S)$.
The matrix $G(\alpha)$ is defined block-wise by 
\[
G(\alpha)\elem[\cP^\mu, \cU] = G^\mu(\alpha) = \sum\nolimits_{S\subseteq [n]} \alpha(\mu, S)\Psi^\mu_S.
\]
Eq.~\rf{eqn:PsiDelta} gives the following important relation:
\begin{equation}
\label{eqn:GDelta}
G(\alpha) \smap{\Delta_j} G(\partial_j\alpha).
\end{equation}

\paragraph{Extended Matrices}
Eq.~\rf{eqn:GDelta} gives one connection between $G(\alpha)$ and the optimisation problem in~\rf{eqn:learningDual}.
Here we give another one.
For that, we define an extended version $\tG(\alpha)$ of $G(\alpha)$.

Let $\ctU = M_\qq\times \cU$.  We use $\ctU^\mu$ to denote $\{\mu\}\otimes \cU$.  The $\ctU\times\cU$-matrix $\tG(\alpha)$ is defined block-wise:
\[
\tG(\alpha)\elem[\cU^\mu, \cU] = \tG^\mu(\alpha) = \sum\nolimits_{S\subseteq [n]} \alpha(\mu, S)\Pi^{[3n]}_S.
\]
Clearly, we have $G(\alpha) = q^{n/2} \tG(\alpha)\elem[\cP,\cU]$.

Using that $\sfigA{\Pi^{[3n]}_S}$ is a decomposition of $\cH^{3n}$ into orthogonal subspaces, we get
\[
\tG(\alpha)^*\tG(\alpha)
= \sum_{\mu\in M_\qq} \sA[\tG^\mu(\alpha)]^* \tG^\mu(\alpha)
= \sum_{S\subseteq [3n]} \skC[ \sum_{\mu\in M_\qq} \alpha(\mu,S)^2 ] \Pi^{[3n]}_S.
\]
As $\|A\| = \sqrt{\|A^*A\|}$ for any matrix $A$, we obtain another important relation:
\begin{equation}
\label{eqn:tGNorm}
\|\tG(\alpha)\| = \|\alpha\|.
\end{equation}
Of course it also holds for $\partial_j\alpha$.
If $\|\tG(\alpha)\|$ and $\|G(\alpha)\|$ were close, then any feasible solution $\alpha$ would give an adversary matrix $\Gamma = G(\alpha)\elem[\cP, \cN]$ with value $\|\alpha\|$.
It is easy to lower bound $\|\Gamma\|$ in terms of $\|\alpha\|$, see \rf{lem:GammaNorm} below, but, in general, $\|G(\partial_j \alpha)\|$ will be much larger than $\|\tG(\partial_j \alpha)\|$.
In particular, this is the case when $\alpha$ is the solution from \rf{prp:learningBoth}.
Our main  challenge in the coming sections will be to find ways to reduce $\|G(\partial_j\alpha)\|$.

\paragraph{Reducing Extended Matrices}
Here we will give a finer relation between $G(\alpha)$ and $\tG(\alpha)$ than the trivial $G(\alpha) = q^{n/2} \tG(\alpha)\elem[\cP,\cU]$. 
For $\cM_\mu \in \cC_\qq$, we define
\begin{equation}
\label{eqn:Psimu}
\Pi^\mu = \sum_{S\subseteq [3n], S\notin \cM_\mu} \Pi^{[3n]}_S
\qqand
\Psi^\mu = \sum_{S\subseteq [3n], S\notin \cM_\mu} \Psi^\mu_S.
\end{equation}
By condition~\rf{eqn:alphaZero}, we have $\tG^\mu(\alpha) = \Pi^\mu \tG^\mu(\alpha)$, and, thus,
$
G^\mu(\alpha) = \Psi^\mu \tG^\mu(\alpha).
$
If we define a linear operator $\Psi_\qq\colon \cH^{\ctU}\to \cH^{\cP}$ by $\Psi_\qq = \bigoplus_{\mu\in M_\qq} \Psi^\mu$, we get
\begin{equation}
\label{eqn:PsiG}
G(\alpha) = \Psi_\qq \tG(\alpha).
\end{equation}
In the light of discussion after~\rf{eqn:tGNorm}, it would help if we could upper bound the norm of $\Psi_\qq$.  Unfortunately, its norm is exponential.
Indeed, we can write
\begin{equation}
\label{eqn:PsimuAlternative}
\Psi^\mu = \bigotimes\nolimits_{T\in\mu} \Psi^T_{\le 2},
\end{equation}
where $\Psi^T_{\le 2} = \sum_{R\subset T, R\ne T} \Psi^T_R$.  We prove its basic properties in the next claim, where we also study the operator $\Psi^T_{\le 1} = \sum_{R\subset T, |R|\le 1} \Psi^T_R$.

\begin{clm}
\label{clm:Psi2Norm}
We have the following estimates
\begin{itemize}
\item[(a)] $\normA|\Psi^T_{\le 2}|=\sqrt{3}$,
\item[(b)] $\normA|\Psi^T_{\le 2}(\Pi_0\otimes I_\cH\otimes I_\cH)| = \normA|\Psi^T_{\le 2}(I_\cH\otimes \Pi_0\otimes I_\cH)| = \normA|\Psi^T_{\le 2}(I_\cH\otimes I_\cH\otimes \Pi_0)|=1$,
\item[(c)] $\normA|\Psi^T_{\le 1}|=1$, 
\item[(d)] $(\Psi^T_\emptyset)^*\Psi^T_{\le 2} = \Pi^T_\emptyset$, and $\|\Psi^T_\emptyset\|=1$.
\end{itemize}
\end{clm}

In the next sections, we will use points (b) and (c) of this lemma to upper bound the norm of $G(\partial_j\alpha)$ using~\rf{eqn:PsiG}.

\pfstart[Proof of \rf{clm:Psi2Norm}]
Let $T=\{t_1,t_2,t_3\}$.
The required identities will easily follow once we establish singular value decompositions of the matrices $\Psi^T_R$.
We identify the triples $(a,b,c)$ in $P^T$ from~\rf{eqn:P} with tuples $(b,c)\in [q]^2$ because $a \equiv -b-c\mod q $ is uniquely determined by $b$ and $c$.
Thus, further in the proof we treat $\Psi^T_R$ as an operator $\Psi^T_R\colon \cH^3\to\cH^2$.

Recall \rf{eqn:PsiTR}, the definition of $\Psi^T_R$ via $\Pi^T_R$.
We can write 
\[
\Pi^T_R = \sum\nolimits_{i_1,i_2,i_3} (e_{i_1}\otimes e_{i_2}\otimes e_{i_3})(e_{i_1}\otimes e_{i_2}\otimes e_{i_3})^*,
\]
where in the sum $i_j$ ranges over $[1..q-1]$ for $t_j\in R$ and $i_j=0$ for $t_j\notin R$. Observe that
\[
\sqrt{q}(e_{i_1}\otimes e_{i_2}\otimes e_{i_3})\elem[-b-c,b,c]
=\omega^{i_1(-b-c)} \omega^{i_2b} \omega^{i_3c} / q
= \omega^{(i_2-i_1)b}\omega^{(i_3-i_1)c} / q
= (e_{i_2-i_1}\otimes e_{i_3-i_1})\elem[b,c].
\]
Thus, $\Psi^T_R$ maps Fourier basis vectors of $\cH^3$ to Fourier basis vectors of $\cH^2$ in accordance to the following mapping
\[
\psi\colon[0..q-1]^3\rightarrow [0..q-1]^2\colon(i_1,i_2,i_3)\mapsto(i_2-i_1,i_3-i_1)\mod q.
\]
In other words, in Fourier basis, any of the matrices in (a), (b), or (c) has exactly one 1 in each column, and its norm is equal to the square root of the maximal number of 1's in a row, which is the number of pre-images of a pair $(i,i')\in[0..q-1]^2$ under $\psi$ subject to restrictions on which of $(i_1,i_2,i_3)$ are allowed to be non-zero.

To prove (a), it now suffices to show that every tuple $(i,i')\in[0..q-1]^2$ has at most three pre-images under $\psi$ of Hamming weight at most $2$. And, indeed, the set of such pre-images is $\{(0,i,i'), (-i,0,i'-i), (-i',i-i',0)\}$.
Note that the cardinality of this set and that of $\{0,i,i'\}$ are equal.
In particular, there is exactly one triple is this set that has $0$ in any given position, which, in turn, proves (b).

We also note that every tuple $(i,i')\in[0..q-1]^2$ has at most one pre-image under $\psi$ of Hamming weight at most $1$ (exactly one pre-image if $|\{0,i,i'\}|\le 2$ and none otherwise). This proves (c).
Finally, the only pre-image of $(0,0)\in[0..q-1]^2$ of Hamming weight at most 2 is $(0,0,0)$, which proves (d) because $\Psi^T_\emptyset = (e_0\otimes e_0)(e_0\otimes e_0\otimes e_0)^*$.
\pfend

\paragraph{Restricting from $\cU$ to $\cN$}
Finally, we give a general way of bounding the norm of $\Gamma = G(\alpha)\elem[\cP,\cN]$ in terms of $\alpha$.  For our upcoming application in~\rf{sec:3matching}, we prove a slightly more general result.  Note that the bound is related to the objective value~\rf{eqn:alphaObjective} of $\alpha$.

\begin{lem}
\label{lem:GammaNorm}
Let $\alpha$ be a solution to the dual learning graph of $\cC_\qq$, and $V$ is an arbitrary linear operator in $\bC^\cU$ satisfying $\Pi^{[3n]}_\emptyset V = \Pi^{[3n]}_\emptyset$.  Then, 
\[
\normB|(G(\alpha)V)\elem[\cP, \cN]| \ge \sqrt{\frac{|\cN|}{|\cU|}\sum\nolimits_{\mu\in M_\qq} \alpha(\mu,\emptyset)^2}.
\]
\end{lem}

\pfstart
Let $\Xi$ be the canonical projector from $\bC^\cU$ to $\bC^\cN$.  Then, $(G(\alpha)V)\elem[\cP, \cN] = G(\alpha)V\Xi^*$.

Fix $\mu$.  We have (we use \rf{clm:Psi2Norm}(d) in the third equality):
\[
(\Psi^\mu_\emptyset)^* G^\mu(\alpha) 
= (\Psi^\mu_\emptyset)^* \Psi^\mu \tG^\mu(\alpha) 
= \sB[\bigotimes\nolimits_{T\in \mu} (\Psi^T_\emptyset)^*\Psi^T_{\le 2}] \tG^\mu(\alpha) 
= \Pi^{[3n]}_\emptyset\tG^\mu(\alpha) 
= \alpha(\mu, \emptyset) \Pi^{[3n]}_\emptyset.
\]
Let $u = e_0^{\otimes 3n}$ be the uniform vector in $\bC^\cU$ and $v$ be the uniform vector in $\bC^\cN$ given by $v\elem[y]=1/\sqrt{|\cN|}$ for each $y\in\cN$.  Then, we have $\Pi^{[3n]}_\emptyset = uu^*$, and, continuing the last equation:
\begin{equation}
\label{eqn:GammaNorm1}
(\Psi^\mu_\emptyset)^* G^\mu(\alpha) V\Xi^*
= \alpha(\mu, \emptyset) \Pi^{[3n]}_\emptyset V\Xi^*  
= \alpha(\mu, \emptyset) \Pi^{[3n]}_\emptyset \Xi^*
= \sqrt{\frac{|\cN|}{|\cU|}}\alpha(\mu, \emptyset) uv^*.
\end{equation}
Let
\begin{equation}
\label{eqn:GammaNormtGamma}
\tGamma = \sB[\bigoplus_{\mu\in M\qq} \Psi^\mu_\emptyset]^*G(\alpha)V\Xi^*.
\end{equation}
This is an $\ctU\times \cN$-matrix.  Its block $\tGamma\elem[\ctU^\mu, \cN]$ is given by~\rf{eqn:GammaNorm1}.  Thus, we have
\[
\tGamma^*\tGamma = \sB[\frac{|\cN|}{|\cU|}\sum\nolimits_{\mu\in M_\qq} \alpha(\mu,\emptyset)^2 ]vv^*.
\]
Hence, using~\rf{eqn:GammaNormtGamma} and \rf{clm:Psi2Norm}(d):
\[
\sqrt{\frac{|\cN|}{|\cU|}\sum\nolimits_{\mu\in M_\qq} \alpha(\mu,\emptyset)^2}
= \normA|\tGamma| 
\le \normC|\bigoplus_{\mu\in M\qq} \Psi^\mu_\emptyset| \norm|G(\alpha)V\Xi^*|
= \normB|(G(\alpha)V)\elem[\cP, \cN]|.\qedhere
\]
\pfend

\section{Lower Bound for the 3-Shift-Sum Problem}
\label{sec:3shift}
The goal is to prove a quantum query lower bound for the 3-shift-sum problem.

\begin{thm}
\label{thm:3shiftLower}
Assume $q\ge 2n^3$.  Then the quantum query complexity of the 3-shift-sum problem is $\Omega(n^{1/3})$.
\end{thm}

The main idea behind the lower bound is to use \rf{clm:Psi2Norm}(c).
In order to do that, we perform a transition to a different certificate structure $\cC'_\sss$.
For each $\mu\in M_\sss$, there is a corresponding $\cM'_\mu \in \cC_\sss'$ obtained as follows: a subset $S\subseteq[3n]$ is in $\cM'_\mu$ if and only if there exists a triple $T\in \mu$ satisfying $|T\cap S|\ge 2$.
Note that this is \emph{not} the certificate structure for the 3-shift-sum problem.
Rather it is the certificate structure of a problem one might call the 3-shift-equal problem.  The input is a $3\times n$-matrix.  In the positive case, there exist circular shifts of rows such that the elements in each column become equal.  In the negative case, any two elements from two different rows are different.

\begin{prp}
\label{prp:learningShift}
The learning graph complexity of the certificate structure $\cC'_\sss$ is $\Omega(n^{1/3})$.
\end{prp}

\pfstart
The proof is similar to that of Proposition 12 from~\cite{belovs:onThePower} for the hidden shift problem.
We have $|\cC_\sss'|=n^2$.
Define
\begin{equation}
\label{eqn:alphacMS}
\alpha(\cM, S) = \frac1n \max\sfig{n^{1/3} - |S|, 0}\qquad \text{if $S\notin \cM$},
\end{equation}
and as 0 otherwise.
It is easy to see that the objective value~\rf{eqn:alphaObjective} is $n^{1/3}$, and that~\rf{eqn:alphaZero} holds.

Fix $S$ and $j$, and let us check~\rf{eqn:alphaOne}.
If $|S|\ge n^{1/3}$, then the left-hand side of~\rf{eqn:alphaOne} is zero, so assume $|S|\le n^{1/3}$.
There are $n^2$ choices of $\cM\in \cC_\sss'$.  
If $S\cup\{j\}\notin\cM$, then the value of $\alpha(\cM, S)$ changes by $1/n$ as the size of $S$ increases by 1.
Also, there are at most $|S|n \le n^{4/3}$ choices of $\cM$ such that $S\notin\cM$ but $S\cup\{j\}\in\cM$.  For each of them, the value of $\alpha(\cM, S)$ changes by at most $n^{-2/3}$.  Thus,
\[
\sum_{\cM\in \cC} \sA[\alpha(\cM, S) - \alpha(\cM, S\cup\{j\})]^2
\le n^2\cdot \frac1{n^2} + n^{4/3}\cdot n^{-4/3} = O(1). \qedhere
\]
\pfend

\subsection{Regular Version}
\label{sec:3shiftRegular}
In this section we prove \rf{thm:3shiftLower}.  Let $\alpha'_\sss$ be the feasible solution~\rf{eqn:alphacMS} for the $\cC'_\sss$ certificate structure.  It is also a feasible solution for the $\cC_\sss$ certificate structure.
As in \rf{sec:basic}, we define the adversary matrix by
\[
\Gamma = G(\alpha'_\sss)\elem[\cP,\cN].
\]
By \rf{lem:GammaNorm}, we get $\|\Gamma\|=\Omega(n^{1/3})$ if we prove that $|\cN| = \Omega(|\cU|)$.
But that is easy: for a uniformly random triple $(a,b,c)\in[q]^3$, the probability that $a+b+c$ is divisible by $q$ is $1/q$.  There are $n^3$ possible triples having one element in each of $A$, $B$, and $C$.  Hence, by the union bound, a uniformly random input in $[q]^{3n}$ is negative with probability at least $1-n^3/q \ge 1/2$.  That is, $|\cN|\ge q^{3n}/2$.

Now let us prove that $\|\Gamma\circ\Delta_j\| = O(1)$.
By~\rf{eqn:GDelta} and using that $\Gamma$ is a submatrix of $G(\alpha'_\sss)$, it suffices to prove that $\|G(\partial_j \alpha'_\sss)\| = O(1)$.

Following~\rf{eqn:Psimu}, let us define an analogue of $\Psi_\sss$ for our new certificate structure $\cC_\sss'$ by
\[
\Psi'^\mu = \sum_{S\subseteq [3n], S\notin \cM'_\mu} \Psi^\mu_S
\qqand
\Psi'_\sss = \bigoplus_{\mu\in M_\sss} \Psi'^\mu.
\]
Similarly to~\rf{eqn:PsiG}, we get
\[
G(\partial_j \alpha'_\sss) = \Psi'_\sss \tG(\partial_j \alpha'_\sss).
\]
We have $\|\tG(\partial_j\alpha'_\sss)\|=O(1)$ by~\rf{eqn:tGNorm} and \rf{prp:learningShift}.  It suffices to prove that $\|\Psi'_\sss\|= O(1)$.  But it is easy to see that 
\[
\Psi'^\mu = \bigotimes\nolimits_{T\in\mu} \Psi^T_{\le 1},
\]
and, by \rf{clm:Psi2Norm}, $\|\Psi'^\mu\|= 1$, hence, $\|\Psi'_\sss\|=1$.

\subsection{Property Testing Version}
\label{sec:property}
In this section, we prove a quantum lower bound for the property testing version of the 3-shift-sum problem.
Unlike the original version of the 3-shift-sum problem, this problem makes sense even for $q=2$, so, for concreteness, we will define it for Boolean alphabet, however, similar results also hold for larger alphabet sizes.

An input is a string in $\bool^{3n}$.
For a positive input $x$, there exists $\mu\in M_\sss$ such that $x_a\oplus x_b\oplus x_c = 0$ for every triple $\{a,b,c\}\in \mu$.  Here $\oplus$ stands for xor.
The negative inputs are defined as being at relative Hamming distance at least $\eps$ to the set of positive inputs.

\begin{thm}
For $\eps \le \frac 17$, the property testing version of the 3-shift-sum problem requires $\Omega(n^{1/3})$ quantum queries to solve.
\end{thm}

The construction is identical to that in \rf{sec:3shiftRegular}.
The proof of $\|\Gamma\circ\Delta_j\| = O(1)$ is identical.  In this part of the proof only $\cP$ and $\cU$ are used, which are the same in the regular and the property testing versions of the problem, and the size of the alphabet is never used.

The only place where the size of the alphabet is used is in lower bounding $\|\Gamma\|$, where it is proven that $|\cN| = \Omega(|\cU|)$.
If we prove this for this version of the problem, we will be done.

Recall that we treat $x$ as an $3\times n$-matrix.
Fix the last two rows.  The input $x$ is negative if its first row is at relative Hamming distance at least $\frac37$ from the xor of any of $n^2$ circular shifts of the last two rows.
A simple application of the Chernoff and the union bounds shows that this is the case with probability $1-o(1)$.

\section{Lower Bound for the 3-Matching-Sum Problem}
\label{sec:3matching}

The goal of this section is to prove the following theorem:

\begin{thm}
\label{thm:3matchingLower}
Assume $q\ge 2n^3$.  Then the quantum query complexity of the 3-matching-sum problem is $\Omega(\sqrt n)$.
\end{thm}

Let $\alpha_\mm$ be the feasible solution~\rf{eqn:alphacMSopt} to the dual learning graph of $\cC_\mm$ from \rf{prp:learningBoth}.  We will obtain an adversary matrix to the 3-matching-sum problem multiplying $G(\alpha_\mm)$ by a suitably chosen projector $V$.
We define it using symmetries of the problem.

The group $\gS_n$ acts on the set $[q]^n$ in the natural way:
$\pi\in\gS_n$ maps $x = (x_1,\ldots,x_n)$ to $\pi x = (x_{\pi^{-1}(1)},\ldots,x_{\pi^{-1}(n)})$, and by linearity we extend this action to $\cH^n$, the latter thus becoming an $\gS_n$-module.

Similarly, the group $\gG=\gS_A\times\gS_B\times\gS_C$ acts on $\cU$: a group element $(\pi_A,\pi_B,\pi_C)\in\gG$ acts on $x = (x_A, x_B, x_C)\in \cU$ by mapping it to $(\pi_Ax_A, \pi_Bx_B, \pi_Cx_C)$.  This action renders $\bC^\cU$ a $\gG$-module.
Let $\gG$ act on $\mu\in M_\mm$ by mapping each triple $(a_1,a_2,a_3)\in\mu$ to $(\pi_A(a_1),\pi_B(a_2),\pi_C(a_3))$.
Together with the action on inputs of length $3n$, this gives us an action of $\gG$ on $\cP$, hence, $\bC^{\cP}$ is also a $\gG$-module.

The 3-matching-sum problem is invariant under this action of $\gG$: positive inputs are mapped to positive inputs, and negative inputs are mapped to negative inputs.  This means that $\bC^\cN$ is a $\gG$-submodule of $\bC^\cU$.  
It is easy to see that $\alpha_\mm$ is symmetric with respect to $\gG$, hence, $G(\alpha_\mm)$ is symmetric with respect to $\gG$.  In other words, it commutes with any element of $\gG$, or, different still, it is a $\gG$-morphism.

Let $T$ be a finite set.
It is easy to see that $\Pi^T_k$ is a $\gS_T$-submodule of $\Pi^T$.
From \cite{belovs:setEquality}, the module $\Pi_k^{T}$ only contains irreps with at most $k$ boxes below the first row.
Denote by $\bar \Pi_k^{T}$ the projector onto the span of all irreps with \emph{exactly} $k$ boxes below the first row.
In particular, $\bar\Pi^T_0 = \Pi^T_0$.

In order to simplify statements of some results, in particular \rf{lem:crazy}, let us assume there is a cutting point $K$ such that
\begin{equation}
\label{eqn:K}
\alpha(\mu, S)=0\quad \text{ whenever } |S|>K.
\end{equation}
For $\alpha_\mm$, we take $K=\floor[\sqrt n]$.  Define the projectors
\[
V^T = \sum_{k = 0}^{K} \bar\Pi^T_k
\qqand
V = V^A\otimes V^B\otimes V^C.
\]
Note that $(\Pi_{k_A}\otimes \Pi_{k_B}\otimes \Pi_{k_C})V = (\bar\Pi_{k_A}\otimes \bar\Pi_{k_B}\otimes \bar\Pi_{k_C})$ for all $k_A, k_B, k_C$ between 0 and $K$.

The adversary matrix $\Gamma$ is obtained as
\[
\Gamma = \sA[G(\alpha_\mm) V]\elem[\cP, \cN].
\]

We know from~\rf{sec:3shiftRegular} that $|\cN| = \Omega(|\cU|)$.  Also $\Pi^{[3n]}_\emptyset V = \bar\Pi^{[3n]}_\emptyset = \Pi^{[3n]}_\emptyset$.  Hence, by \rf{lem:GammaNorm} and \rf{prp:learningBoth}, we have
\[
\|\Gamma\| = \Omega(\sqrt n).
\]

It remains to prove that $\|\Delta_j\circ \Gamma\|=O(1)$ for any $j$, which we do in the remaining part of this section.
Due to symmetry, $\|\Delta_j\circ\Gamma\|$ is the same for all $j$, so it suffices to consider $j=1$.
Note that $\Delta_1\circ\Gamma$ is a $\gG'$-morphism, where $\gG' = \gS_{[2..n]}\otimes \gS_{B}\otimes\gS_{C}$.

We have to understand how $\Delta_1$ acts on $V$, or, in particular, how it acts on $\bar\Pi^{[n]}_k$.  For the usual projector, $\Pi^{[n]}_k$, we have the identity 
$
\Pi^{[n]}_k = \Pi_0\otimes \Pi^{[2..n]}_k + \Pi_1\otimes \Pi^{[2..n]}_{k-1}
$.
Ref.~\cite{belovs:setEquality} gives the following analogue of this identity for $\bar\Pi_k^{[n]}$, where one should think of $\Phi^{[n]}_k$ as an error term.
\begin{lem}
\label{lem:Phik}
Let
\[
\Phi_k^{[n]} = \bar \Pi_k^{[n]} - \Pi_0\otimes \bar \Pi^{[2..n]}_k - \Pi_1\otimes \bar \Pi^{[2..n]}_{k-1}.
\]
If $k < n/3$, then
$\normA|\Phi_k^{[n]}| = O(1/\sqrt{n})$.  
\end{lem}

Define $\Phi^A = \sum_{k=1}^{K} \Phi^A_k$.  It is easy to see that $\Pi^A_k \Phi^A_k\Pi^A_k = \Phi^A_k$, hence, $\|\Phi^A_k\| = O(1/\sqrt{n})$.
Let 
\[
\Phi=\Phi^A\otimes V^B\otimes V^C
\qqand
V' = \Pi_0\otimes V^{[2..n]}\otimes V^B\otimes V^C.
\]

From~\rf{lem:Phik}, we get the following variant of relation~\rf{eqn:GDelta}:

\begin{lem}
\label{lem:GVDelta}
Let $\alpha$ be a solution to $\cC_\mm$ satisfying~\rf{eqn:K}.  Then,
\(
G(\alpha) V \smap{\Delta_1} G(\partial_1\alpha) V' + G(\alpha) \Phi.
\)
\end{lem}

\pfstart
It is enough to consider one block $\hG^\mu(\alpha) V$ corresponding to an arbitrary matching $\mu$.
By linearity, it suffices to consider the case $\hG^\mu(\alpha) = \Psi^\mu_S$ for some $S\subseteq[3n]$ of size at most $K$. 
That is, by~\rf{eqn:PsiDelta}, it suffices to prove 
%
\[
\text{$\Psi^\mu_SV \smap{\Delta_1} \Psi_S^\mu V' + \Psi^\mu_S \Phi $\quad if\quad $1\notin S$}
\qqand
\text{$\Psi^\mu_SV \smap{\Delta_1} -\Psi_{S\setminus\{1\}}^\mu V' + \Psi^\mu_S \Phi$\quad if\quad $1\in S$}.
\]

Recall that $\Psi^\mu_S = \Psi^\mu \Pi^{[3n]}_S$, and we have
\begin{equation}
\label{eqn:GammaV0}
\Pi^{[3n]}_S V = 
(\Pi^A_{S\cap A}V^A)\otimes(\Pi^B_{S\cap B}V^B)\otimes(\Pi^C_{S\cap C}V^C).
\end{equation}
We are interested in the first multiplier.  Let $R = S\cap A$ and $k = |R|$.
\begin{equation}
\label{eqn:GammaV1}
\Pi^A_{S\cap A}V^A = \Pi^A_{R}\Pi^A_k V^A = \Pi^A_R \bar\Pi_k^A = \Pi^A_R \sA[\Pi_0\otimes \bar \Pi^{[2..n]}_k + \Pi_1\otimes \bar \Pi^{[2..n]}_{k-1} + \Phi_k^{A}].
\end{equation}
There are two cases.  If $1\notin R$, then~\rf{eqn:GammaV1} is equal to
\[
\Pi^A_R (\Pi_0\otimes \bar\Pi^{[2..n]}_k) + \Pi^A_R \Phi_k^{A} 
\smap{\Delta_1}
\Pi^A_R (\Pi_0\otimes \bar\Pi^{[2..n]}_k) + \Pi^A_R \Phi_k^{A},
\]
and if $1\in R$, it is equal to 
\[
\Pi^A_R (\Pi_1\otimes \bar \Pi^{[2..n]}_{k-1}) + \Pi^A_R \Phi_k^{A} 
\smap{\Delta_1}
-\Pi^A_{R\setminus\{1\}} (\Pi_0\otimes \bar \Pi^{[2..n]}_{k-1}) + \Pi^A_R \Phi_k^{A}.
\]
By plugging it into~\rf{eqn:GammaV0}, and using $\Pi^A_k \Phi^A_k\Pi^A_k = \Phi^A_k$, we get the required identity.
\pfend

Applying \rf{lem:GVDelta} to $G(\alpha_\mm)$, we obtain
\[
G(\alpha_\mm) V \smap{\Delta_1} G(\partial_1\alpha_\mm) V' + G(\alpha_\mm) \Phi.
\]
Denote $W = I^A\otimes V^B\otimes V^C$, where $I^A$ is the identity operator on $\cH^A$.  Note that $V' = WV'$ and $\Phi = W\Phi$.  Also, $\|V'\| = 1$ and $\|\Phi\| = O(1/\sqrt{n})$.  Thus, since $\Gamma$ is a submatrix of $G(\alpha_\mm)V$, it suffices to prove that
\begin{equation}
\label{eqn:nado}
\normA|G(\partial_1\alpha_\mm)W| = O(1)
\qqand
\normA|G(\alpha_\mm)W| = O(\sqrt{n}),
\end{equation}
which is reasonable since $\|\partial_1\alpha_\mm\| = O(1)$ and $\|\alpha_\mm\| = O(\sqrt n)$.
We prove this using the following somewhat technical estimate on the norm of $G(\alpha)W$:

\begin{lem}
\label{lem:crazy}
For $\alpha$ be a solution to the dual learning graph of $\cC_\mm$ satisfying~\rf{eqn:K} and symmetric with respect to $\gS_B\times\gS_C$.
Then
\[
\|G(\alpha)W\|
\le
\max_{k_B,k_C\in [0..K]} \Lambda_{k_B, k_C}(\alpha),
\]
where $\Lambda_{k_B, k_C}(\alpha)$ is defined in the following way.
Let $R_B = [n+1..n+2k_B]$ and $R_C = [2n+1..2n+2k_C]$.
Let $L(\mu, R_B, R_C)$ be the number of triples in the matching $\mu$ that intersect both $R_B$ and $R_C$.
Then
\begin{equation}
\label{eqn:Lambda}
\Lambda_{k_B, k_C}(\alpha) =
\sqrt{\sum_{\mu\in M_\mm} 3^{L(\mu, R_B, R_C)} 
\max_{S\subseteq A\cup R_B\cup R_C}
 \alpha(\mu, S)^2}.
\end{equation}
\end{lem}

This lemma is proven in \rf{sec:crazy}, and now we show how to use it to prove the estimates in~\rf{eqn:nado}.
The exponential term in~\rf{eqn:Lambda} might be somewhat of a concern, but we prove that the fraction of matchings with large $L(\mu, R_B, R_C)$ decreases even faster.

\begin{lem}
\label{lem:L}
Assume $|R_B|, |R_C|\le 2\sqrt n$.  Then, 
$\Pr_\mu \skA[L(\mu, R_B, R_C) = \ell] \le {8^\ell}/{\ell!}$,
where the probability is over uniformly random $\mu\in M_\mm$.
\end{lem}

\pfstart
Fix $\ell$ elements in each $R_B$ and $R_C$.
The probability that these elements are mutually matched by a random $\mu$ is $\binom{n}{\ell}^{-1}$. Hence, by the union bound, the probability that for a randomly chosen $\mu$ there are $\ell$ (or more) elements in $R_B$ matched to elements in $R_C$ is at most
\[
\binom{|R_B|}{\ell} \binom{|R_C|}{\ell} \bigg/ \binom{n}{\ell} \le \frac{(2\sqrt n)^{2\ell}}{\ell!(n/2)^\ell}\le\frac{8^\ell}{\ell!},
\]
where we have assumed that $n$ is large enough so that $\ell\le 2\sqrt n < n/2$.
\pfend

\begin{clm}
We have $\normA|G(\alpha_\mm)W| = O(\sqrt n)$.
\end{clm}

\pfstart
We apply \rf{lem:crazy}.  By~\rf{eqn:alphacMSopt}, we have $\alpha_\mm(\mu,S)^2 \le  n/|M_\mm|$ for all $\mu$ and $S$.  Hence,
\[
\Lambda_{k_B, k_C}(\alpha_\mm) \le \sqrt{n}\sqrt{\bE_{\mu\in M_\mm} \skA[3^{L(\mu, R_B, R_C)}]}.
\]
And, using \rf{lem:L}:
\begin{equation}
\label{eqn:LambdaOcenka}
\bE_{\mu\in M_\mm} \skA[3^{L(\mu, R_B, R_C)}]
\le
\sum_{\ell=0}^{\infty} 3^\ell\cdot \frac{8^\ell}{\ell!} = \ee^{24} = O(1),
\end{equation}
which gives the required bound.
\pfend

\begin{clm}
We have $\normA|G(\partial_1\alpha_\mm)W| = O(1)$.
\end{clm}

\pfstart
We apply \rf{lem:crazy}.
By the analysis in \rf{prp:learningBoth}, we see that 
\[
\max_{S\subseteq A\cup R_B\cup R_C}
 \partial_1\alpha_\mm(\mu, S)^2 \le
\frac{1}{|M_\mm|}
\begin{cases}
n, &\text{if 1 is matched by $\mu$ to elements in both $R_B$ and $R_C$;}\\
1, &\text{otherwise.}
\end{cases}
\]
Let us call the event in the first case above $Z(\mu)$.
Then,
\[
\Lambda_{k_A, k_B}(\partial_1\alpha_\mm)^2 \le
\bE_{\mu\in M_\mm} \sk[ 3^{L(\mu, R_B, R_C)}] 
+ \Pr\nolimits_\mu[Z(\mu)] \cdot 3n\, \bE_{\mu\in M_\mm} \sk[ 3^{L(\mu, R_B, R_C)-1} \midB Z(\mu) ].
\]
The first term is $O(1)$ by~\rf{eqn:LambdaOcenka}.  For the second term, it is easy to see that $\Pr_\mu[Z(\mu)] = |R_B||R_C|/n^2 = O(1/n)$, and the conditioned expectation the same as in~\rf{eqn:LambdaOcenka}, because we can remove the triple containing 1 from consideration thus reducing to the same problem with $n$, $|R_B|$, and $|R_C|$ smaller by 1.
This gives the required bound of $O(1)$.
\pfend

\subsection{Proof of \rf{lem:crazy}}
\label{sec:crazy}

We need the following technical result from~\cite{belovs:setEquality}.
Let $F$ be the element of the group algebra $\bC\gS_n$ defined by
\begin{equation}
\label{eqn:kappa}
F = \frac{1}{2^{k}}\; \sA[\eps - (a_1,b_1)]\sA[\eps - (a_2,b_2)]\cdots \sA[\eps - (a_k, b_k)],
\end{equation}
where $a_1,b_1,\dots,a_k,b_k$ are some distinct elements of $[n]$, $\eps$ is the identity element of $\gS_n$, and $(a_i,b_i)$ denotes the transposition of $a_i$ and $b_i$.
Note that $F$ is an orthogonal projector.

\begin{lem}[\cite{belovs:setEquality}]
\label{lem:kappa}
Consider an irrep $S^\lambda$ of $\gS_n$ with exactly $k$ boxes below the first row.
There exists a non-zero vector $v\in S^\lambda$ such that $F v = v$, where $F$ is defined in~\rf{eqn:kappa}.
\end{lem}

Now we start with the proof.
Since $\alpha$ is symmetric with respect to $\gS_B\times\gS_C$, 
the operator $G(\alpha)W$ is an $\gS_B\times \gS_C$-morphism.
By \rf{lem:singularirrep}, there exists a copy $\cS$ of an $\gS_B\times \gS_C$-irrep $S^{\lambda_B}\otimes S^{\lambda_C}$ consisting of principal right-singular vectors of $G(\alpha)W$.
Let $k_B$ and $k_C$ be the number of boxes below the first row in $\lambda_B$ and $\lambda_C$, respectively.
By the construction of $V^B$ and $V^C$, $k_B, k_C\le K$ and this irrep $\cS$ belongs to $I^A\otimes \bar\Pi^B_{k_B} \otimes\bar\Pi^C_{k_C} \subseteq I^A\otimes \Pi^B_{k_B} \otimes \Pi^C_{k_C}$.

Applying \rf{lem:kappa} to both $S^{\lambda_B}$ and $S^{\lambda_C}$ and taking the product of the results, we get a normalised principal right-singular vector $v$ of $G(\alpha)W$ satisfying $Fv = v$, where $F$ is defined in~\rf{eqn:kappa}.  Here $k = k_A + k_B$, and, out of $k$ pairs $\{a_i,b_i\}$, $k_B$ pairs are contained in $B$ and $k_C$ pairs are contained in $C$.
For concreteness, let us take $\{a_i,b_i\} = \{n+2i-1,n+2i\}$ for $i\le k_B$ and $\{a_i,b_i\} = \{2n+2i-2k_B-1, 2n+2i-2k_B\}$ for $i>k_B$.
Let us denote $R_B = [n+1..n+2k_B]$ and $R_C = [2n+1..2n+2k_C]$ so that $R_B\cup R_C$ is the set of all $a_i$ and $b_i$.

Consider the vector $v$ in the Fourier basis.  Each basis vector is nullified by $F$ unless there is at least one non-zero component in each of $\{a_i,b_i\}$.  Since $v$ is in $I^A\otimes \Pi^B_{k_B} \otimes \Pi^C_{k_C}$, there are exactly $k_B$ and $k_C$ non-zero components in $B$ and $C$ respectively.  Hence, they are all contained in $R_B\cup R_C$.  The above discussion implies that $v$ belongs to $\Xi$, where
\[
\Xi = I^A\otimes I^{R_B\cup R_C} \otimes \Pi^{B\cup C\setminus(R_B\cup R_C)}_0.
\]
We have
\[
\normA|G(\alpha)W| = \normA|G(\alpha)Wv| = \normA|G(\alpha)v| = \normA|G(\alpha)\Xi v|\le \normA|G(\alpha)\Xi|.
\]
In particular,
\[
\norm|G(\alpha)W|^2 
\le \norm|\sA[G(\alpha)\Xi]^* G(\alpha)\Xi | 
\le \sum_{\mu\in M_\mm} \normA|G^\mu(\alpha)\Xi|^2.
\]
Fix $\mu\in M_\mm$.  In the block corresponding to $\mu$, we have
\[
\normA|G^\mu(\alpha)\Xi| = \normA|\Psi^\mu \tG^\mu(\alpha)\Xi| \le \|\Psi^\mu\Xi\|\, \normA|\tG^\mu(\alpha)\Xi|.
\]
For the first multiplier, we have $\|\Psi^\mu\Xi\| \le 3^{L(\mu, R_B, R_C)/2}$ by~\rf{eqn:PsimuAlternative} and \rf{clm:Psi2Norm} points (a) and (b).  For the second multiplier, we have
\[
\normA|\tG^\mu(\alpha)\Xi| = \max_{S\subseteq A\cup R_B\cup R_C} \alpha(\mu, S),
\]
and the lemma follows.


\section{Open Problems}
The obvious open problem is to resolve the quantum query complexity of the 3-shift-sum problem.  So far we only have an $\Omega(n^{1/3})$ lower bound and an $O(n^{1/2})$ upper bound, however, it is not clear how to improve on either of them.

For the 3-matching-sum problem, we have proven matching upper and lower bounds of $\Theta(\sqrt n)$.  An interesting problem is to generalise this to the $k$-matching-sum problem for arbitrary $k$.  The main limitation seems to be the norm of the error term in \rf{lem:Phik}. 

Some other open problems can be formulated.
What functions with randomised query complexity $\omega(\sqrt n)$ could potentially have poly-logarithmic quantum query complexity?  Or, can a relatively general result be proven that excludes some of such functions?
For what other problems can the dual learning graph framework be useful?

\section*{Acknowledgements}
This research is partially supported by the ERDF project number 1.1.1.2/I/16/113.
Part of this work was done while supported by the ERC Advanced Grant MQC.
This research is partially supported by the Singapore Ministry of Education and the National Research Foundation.

Part of this research was done while A.B. was visiting the Centre for Quantum Technologies at the National University of Singapore.  A.B. would like to thank Miklos Santha for hospitality.

\small
\bibliographystyle{habbrvM}
\bibliography{belov}

\begin{thebibliography}{10}

\bibitem{aaronson:forrelation}
S.~Aaronson and A.~Ambainis.
\newblock Forrelation: A problem that optimally separates quantum from
  classical computing.
\newblock In {\em Proc.\ of 47th ACM STOC}, pages
  \myhref{http://dx.doi.org/10.1145/2746539.2746547}{307--316}, 2015.
\newblock  \myhref{http://arxiv.org/abs/1411.5729}{{\ttfamily
  arXiv:1411.5729}}.

\bibitem{aaronson:cheatSheets}
S.~Aaronson, S.~Ben-David, and R.~Kothari.
\newblock Separations in query complexity using cheat sheets.
\newblock In {\em Proc.\ of 48th ACM STOC}, pages
  \myhref{http://dx.doi.org/10.1145/2897518.2897644}{863--876}, 2016.
\newblock  \myhref{http://arxiv.org/abs/1511.01937}{{\ttfamily
  arXiv:1511.01937}}.

\bibitem{ambainis:adv}
A.~Ambainis.
\newblock Quantum lower bounds by quantum arguments.
\newblock {\em Journal of Computer and System Sciences},
  64(4):\myhref{http://dx.doi.org/10.1006/jcss.2002.1826}{750--767}, 2002.
\newblock Earlier: \myhref{http://dx.doi.org/10.1145/335305.335394}{{\em
  STOC'00}},  \myhref{http://arxiv.org/abs/quant-ph/0002066}{{\ttfamily
  arXiv:quant-ph/0002066}}.

\bibitem{belovs:learning}
A.~Belovs.
\newblock Span programs for functions with constant-sized 1-certificates.
\newblock In {\em Proc.\ of 44th ACM STOC}, pages
  \myhref{http://dx.doi.org/10.1145/2213977.2213985}{77--84}, 2012.
\newblock  \myhref{http://arxiv.org/abs/1105.4024}{{\ttfamily
  arXiv:1105.4024}}.

\bibitem{belovs:phd}
A.~Belovs.
\newblock {\em Applications of the Adversary Method in Quantum Query
  Algorithms}.
\newblock PhD thesis, University of Latvia, 2014.
\newblock  \myhref{http://arxiv.org/abs/1402.3858}{{\ttfamily
  arXiv:1402.3858}}.

\bibitem{belovs:setEquality}
A.~Belovs and A.~Rosmanis.
\newblock Adversary lower bounds for the collision and the set equality
  problems.
\newblock  \myhref{http://arxiv.org/abs/1310.5185}{{\ttfamily
  arXiv:1310.5185}}, 2013.

\bibitem{belovs:onThePower}
A.~Belovs and A.~Rosmanis.
\newblock On the power of non-adaptive learning graphs.
\newblock {\em Computational Complexity},
  23(2):\myhref{http://dx.doi.org/10.1007/s00037-014-0084-1}{323--354}, 2014.
\newblock Earlier: \myhref{http://dx.doi.org/10.1109/CCC.2013.14}{{\em
  CCC'13}},  \myhref{http://arxiv.org/abs/1210.3279}{{\ttfamily
  arXiv:1210.3279}}.

\bibitem{spalek:kSumLower}
A.~Belovs and R.~{\v Spalek}.
\newblock Adversary lower bound for the {$k$-sum} problem.
\newblock In {\em Proc.\ of 4th ACM ITCS}, pages
  \myhref{http://dx.doi.org/10.1145/2422436.2422474}{323--328}, 2013.
\newblock  \myhref{http://arxiv.org/abs/1206.6528}{{\ttfamily
  arXiv:1206.6528}}.

\bibitem{ben-david:super-grover}
S.~Ben-David.
\newblock A super-{G}rover separation between randomized and quantum query
  complexities.
\newblock  \myhref{http://arxiv.org/abs/1506.08106}{{\ttfamily
  arXiv:1506.08106}}, 2015.

\bibitem{buhrman:querySurvey}
H.~Buhrman and R.~de~Wolf.
\newblock Complexity measures and decision tree complexity: a survey.
\newblock {\em Theoretical Computer Science},
  288:\myhref{http://dx.doi.org/10.1016/S0304-3975(01)00144-X}{21--43}, 2002.

\bibitem{curtis:representationTheory}
C.~W. Curtis and I.~Reiner.
\newblock {\em Representation theory of finite groups and associative
  algebras}.
\newblock AMS, 1962.

\bibitem{ettinger:hspQuery}
M.~Ettinger, P.~H{\o}yer, and E.~Knill.
\newblock The quantum query complexity of the hidden subgroup problem is
  polynomial.
\newblock {\em Information Processing Letters},
  91(1):\myhref{http://dx.doi.org/10.1016/j.ipl.2004.01.024}{43--48}, 2004.
\newblock  \myhref{http://arxiv.org/abs/quant-ph/0401083}{{\ttfamily
  arXiv:quant-ph/0401083}}.

\bibitem{hoyer:advNegative}
P.~H{\o}yer, T.~Lee, and R.~{\v Spalek}.
\newblock Negative weights make adversaries stronger.
\newblock In {\em Proc.\ of 39th ACM STOC}, pages
  \myhref{http://dx.doi.org/10.1145/1250790.1250867}{526--535}, 2007.
\newblock  \myhref{http://arxiv.org/abs/quant-ph/0611054}{{\ttfamily
  arXiv:quant-ph/0611054}}.

\bibitem{james:symmetricGroup}
G.~James and A.~Kerber.
\newblock {\em The Representation Theory of the Symmetric Group}, volume~16 of
  {\em Encyclopedia of Mathematics and its Applications}.
\newblock Addison-Wesley, 1981.

\bibitem{kuperberg:dihedral}
G.~Kuperberg.
\newblock A subexponential-time quantum algorithm for the dihedral hidden
  subgroup problem.
\newblock {\em SIAM Journal on Computing}, 35:170--188, 2005.
\newblock  \myhref{http://arxiv.org/abs/quant-ph/0302112}{{\ttfamily
  arXiv:quant-ph/0302112}}.

\bibitem{lee:stateConversion}
T.~Lee, R.~Mittal, B.~W. Reichardt, R.~{\v Spalek}, and M.~Szegedy.
\newblock Quantum query complexity of state conversion.
\newblock In {\em Proc.\ of 52nd IEEE FOCS}, pages
  \myhref{http://dx.doi.org/10.1109/FOCS.2011.75}{344--353}, 2011.
\newblock  \myhref{http://arxiv.org/abs/1011.3020}{{\ttfamily
  arXiv:1011.3020}}.

\bibitem{montanaro:quantumProperyTest}
A.~Montanaro and R.~de~Wolf.
\newblock A survey of quantum property testing.
\newblock {\em Theory of Computing Graduate Surveys},
  7:\myhref{http://dx.doi.org/10.4086/toc.gs.2016.007}{1--81}, 2016.
\newblock  \myhref{http://arxiv.org/abs/1310.2035}{{\ttfamily
  arXiv:1310.2035}}.

\bibitem{reichardt:spanPrograms}
B.~W. Reichardt.
\newblock Span programs and quantum query complexity: The general adversary
  bound is nearly tight for every {B}oolean function.
\newblock In {\em Proc.\ of 50th IEEE FOCS}, pages
  \myhref{http://dx.doi.org/10.1109/FOCS.2009.55}{544--551}, 2009.
\newblock  \myhref{http://arxiv.org/abs/0904.2759}{{\ttfamily
  arXiv:0904.2759}}.

\bibitem{sagan:symmetricGroup}
B.~E. Sagan.
\newblock {\em The symmetric group: representations, combinatorial algorithms,
  and symmetric functions}, volume 203 of {\em Graduate Texts in Mathematics}.
\newblock Springer, 2001.

\bibitem{serre:representation}
J.-P. Serre.
\newblock {\em Linear Representations of Finite Groups}, volume~42 of {\em
  Graduate Texts in Mathematics}.
\newblock Springer, 1977.

\bibitem{shi:collisionLowerOriginal}
Y.~Shi.
\newblock Quantum lower bounds for the collision and the element distinctness
  problems.
\newblock In {\em Proc.\ of 43th IEEE FOCS}, pages
  \myhref{http://dx.doi.org/10.1109/SFCS.2002.1181975}{513--519}, 2002.
\newblock  \myhref{http://arxiv.org/abs/quant-ph/0112086}{{\ttfamily
  arXiv:quant-ph/0112086}}.

\bibitem{spalek:advEquivalent}
R.~{\v Spalek} and M.~Szegedy.
\newblock All quantum adversary methods are equivalent.
\newblock {\em Theory of Computing},
  2:\myhref{http://dx.doi.org/10.4086/toc.2006.v002a001}{1--18}, 2006.
\newblock Earlier: \myhref{http://dx.doi.org/10.1007/11523468_105}{{\em
  ICALP'05}},  \myhref{http://arxiv.org/abs/quant-ph/0409116}{{\ttfamily
  arXiv:quant-ph/0409116}}.

\bibitem{zhandry:setEquality}
M.~Zhandry.
\newblock A note on the quantum collision and set equality problems.
\newblock {\em Quantum Information \& Computation}, 15(7\&8):557--567, 2015.
\newblock  \myhref{http://arxiv.org/abs/1312.1027}{{\ttfamily
  arXiv:1312.1027}}.

\bibitem{zhang:advPower}
S.~Zhang.
\newblock On the power of {Ambainis} lower bounds.
\newblock {\em Theoretical Computer Science},
  339(2):\myhref{http://dx.doi.org/10.1016/j.tcs.2005.01.019}{241--256}, 2005.
\newblock  \myhref{http://arxiv.org/abs/quant-ph/0311060}{{\ttfamily
  arXiv:quant-ph/0311060}}.

\end{thebibliography}

\end{document}